# Transfer Learning Strategies for Pathological Foundation Models: A Systematic Evaluation in Brain Tumor Classification

Ken Enda[1], Yoshitaka Oda[1], Zen-ichi Tanei[2], Kenichi Satoh[3], Hiroaki Motegi[4], Terasaka Shunsuke[5], Shigeru Yamaguchi[6], Takahiro Ogawa[7], Wang Lei[1, 8], Masumi Tsuda[1, 8], Shinya Tanaka[1, 2, 8].

[1] Department of Cancer Pathology, Faculty of Medicine, Hokkaido University, Sapporo, Japan.
[2] Department of Surgical Pathology, Hokkaido University Hospital, Sapporo, Japan.
[3] Department of Neurosurgery, Nakamura Memorial Hospital, Sapporo, Japan.
[4] Department of Neurosurgery, Sapporo Azabu Neurosurgical Hospital, Sappro, Japan.
[5] Department of Neurosurgery, Sapporo Hakuyokai Hospital, Sapporo, Japan.
[6] Department of Neurosurgery, Faculty of Medicine, Hokkaido University, Sapporo, Japan.
[7] Faculty of Information Science and Technology, Hokkaido University, Sapporo, Japan.
[8] Institute for Chemical Reaction Design and Discovery (WPI-ICReDD), Hokkaido University, Sapporo, Japan.

## Abstract

Foundation models pretrained on large-scale pathology datasets have shown promising results across various diagnostic tasks. Here, we present a systematic evaluation of transfer learning strategies for brain tumor classification using these models. We analyzed 254 cases comprising five major tumor types: glioblastoma, astrocytoma, oligodendroglioma, primary central nervous system lymphoma, and metastatic tumors. Comparing state-of-the-art foundation models with conventional approaches, we found that foundation models demonstrated robust classification performance with as few as 10 patches per case, despite the traditional assumption that extensive per-case image sampling is necessary. Furthermore, our evaluation revealed that simple transfer learning strategies like linear probing were sufficient, while fine-tuning often degraded model performance. These findings suggest a paradigm shift from "training encoders on extensive pathological data" to "querying pre-trained encoders with labeled datasets", providing practical implications for implementing AI-assisted diagnosis in clinical pathology.

## Corresponding author

Shinya Tanaka
Department of Cancer Pathology, Faculty of Medicine, Hokkaido University
Phone: +81-11-706-5052. 7381, Fax +81-11-706-5902
e-mail: tanaka@med.hokudai.ac.jp
ORCID iD: https://orcid.org/0000-0001-6470-3301

# Introduction

An accurate diagnosis of malignant brain tumors is required for appropriate treatment selection and improved prognosis of the patients [1], however, the decision of pathological diagnosis by morphological observation of the tissue section is difficult even for certified pathologists because of their morphological variety compared with the cancers originating from other organs. The major malignant brain tumors include gliomas, primary central nervous system lymphomas (PCNSL), and metastatic brain tumors, which often present similar contrast enhancement patterns (ring enhancement) on imaging, making preoperative differential diagnosis challenging [2]. According to the Japan Neurosurgical Society's comprehensive survey [3], the distribution of brain tumor surgeries shows: metastatic brain tumors (30.6%), glioblastoma (13.0%), PCNSL (4.9%), astrocytoma (4.7%), and oligodendroglioma (1.9%). These proportions provide insights into real-world diagnostic scenarios and pretest probabilities in clinical settings. Furthermore, as these tumors differ significantly in their treatment strategies and prognosis, accurate differential diagnosis directly impacts patient survival [4].

In the process of pathological diagnosis, immunohistochemical markers and molecular diagnostic techniques are used in addition to morphological observation [5]. The WHO Classification of Central Nervous System Tumors 5th edition [6], published in 2021, introduced diagnostic criteria incorporating molecular pathological features, demanding more precise diagnosis[6]. Particularly for adult-type diffuse gliomas, detection of IDH1/2 (isocitrate dehydrogenase 1/2) gene mutation status and chromosomal 1p/19q codeletion by genetic analysis has become mandatory for diagnosis of glioblastoma, astrocytoma, and oligodendroglioma. However, these molecular methods are time-consuming and require specialized facilities, thus rapid diagnosis cannot be achieved particularly for intraoperative diagnosis to determine the initial treatment.

To solve these issues and for accurate and rapid diagnosis, AI applications in pathological image analysis of brain tumors have gained attention [7], which could potentially provide rapid and cost-effective diagnostic support.In addition, discrimination between glioma and lymphoma is crucial because the therapeutic modality is different as surgical resection is the 1st choice for glioma, but chemotherapy should be done for lymphoma without any surgical damage of the brain parenchyma. In this context, to date, a number of studies have been reported primarily focused on *IDH1/2* gene mutation-based subtyping and grading of gliomas [8–10]. However, fewer studies using AI have been done including metastatic tumors and lymphomas [11].

Following the breakthrough of DINOv2 in self-supervised learning (SSL) [12], specialized foundation models for histopathology emerged, trained on unprecedented scales of diverse pathological images. Several models were introduced in succession, including UNI [13] based on ViT-L(VisionTransformer Large [14]) architecture, Prov-GigaPath [15] based on ViT-Giant architecture, Virchow [16], and Virchow2 [17]. These next-generation foundation models demonstrated significantly superior performance across almost all subtasks compared to conventional benchmarks such as CTransPath[18] and REMEDIS[19].

When applying these foundation models to actual tasks such as classification, various transfer learning strategies have been explored, with fine-tuning and linear probing being two major approaches [20]. While these strategies have been extensively studied for traditional deep learning models, systematic comparisons specifically focusing on pathology foundation models, particularly in brain tumor classification, remain limited.

In this study, we conduct a comprehensive evaluation of transfer learning strategies through three key comparisons (Fig. 1). First, fine-tuning and linear probing approaches using UNI within the context of brain tumor pathology were compared. Second, UNI against standard benchmarks (CTransPath and ImageNet[21] pretrained ViT-L) to assess the impact of pathology-specific pretraining were evaluated. Third, we compared UNI with Prov-GigaPath to evaluate current-generation foundation models of different scales. These comparisons were conducted using both a local malignant brain tumor dataset and an external dataset from EBRAINS [22].

Through this investigation, we aimed to clarify the optimal training strategies for foundation models in real-world clinical settings, with particular focus on their generalization capabilities when applied to external datasets. By comparing various transfer learning approaches and analyzing the impact of training data volume, we searched to provide practical insights for implementing AI-assisted diagnosis in clinical pathology. The findings were expected to advance the development and clinical application of more effective AI models in medical image analysis. Additionally, as a demonstration of practical application, a prototype of web-based system was developed for interactive classification of brain tumor pathology images.

# Results

## Results of Cross-Validation on Local Dataset

In our classification tasks, we evaluated performance across five tumor types: Glioblastoma (G), Astrocytoma (A), Oligodendroglioma (O), primary central nervous system lymphoma (PCNSL, L), and metastatic tumor (M) classes. The complete performance metrics for all models with 500 patches/case are presented (Table 1). Class-wise recall metrics are shown (Supplemental Table 1). Due to consistently poor performance, ResNet-RS 50IN(LP) was excluded from subsequent figures for clarity.

Training with 500 patches/case on the Local dataset, macro recall and patch-wise accuracy for both fine-grained and coarse-grained classifications are shown respectively (Fig. 2a). Foundation models UNI(LP), UNI(FT), and Prov-GigaPath(LP) demonstrated robust performance across all metrics, with no significant differences among them ($p>0.1$). Quantitatively, these models achieved similarly high macro recall (Prov-GigaPath(LP): 0.88±0.10, UNI(FT): 0.87±0.07, UNI(LP): 0.88±0.09) and patch accuracy (Prov-GigaPath(LP): 0.93±0.02, UNI(FT): 0.94±0.01, UNI(LP): 0.93±0.03) in fine-grained classification. Notably, in terms of overall accuracy, Prov-GigaPath(LP) achieved the highest performance in coarse-grained classification (macro recall: 0.99±0.01, patch accuracy: 0.97±0.01), correctly classifying 251/254 cases, while both Prov-GigaPath(LP) and UNI(LP) performed best in fine-grained classification, each correctly classifying 234/254 cases.

UNI(LP) significantly outperformed traditional models with fine-tuned encoders (ViT$_{IN}$(FT) and CTransPath(FT)) in patch-wise accuracy (both $p<0.01$). Notably, only UNI and Prov-GigaPath with Linear Probing exceeded the performance of Fine-tuned models. In contrast, comparing training conditions within the same model architecture, both ViT$_{IN}$ and CTransPath showed higher patch accuracy in FT versus LP conditions (both $p<0.01$). ResNet-RS 50(LP) showed consistently poor performance and was excluded from the graphs due to non-competitive scores. The confusion matrices for UNI(LP) and UNI(FT) revealed frequent misclassification of Astrocytoma cases as either Glioblastoma or Oligodendroglioma.

The confusion matrices for the Local dataset (Fig. 2c) showed high classification accuracy with similar patterns between LP and FT. Both conditions demonstrated accurate classification of metastatic tumors and PCNSL, with misclassifications primarily occurring within glioma classes.

## Results on External Dataset

Macro recall and patch accuracy results on the EBRAINS dataset are presented (Fig. 2b). UNI(LP) demonstrated superior performance across all metrics (macro recall: 0.75±0.01, patch accuracy: 0.65±0.02 for fine-grained; macro recall: 0.94±0.01, patch accuracy: 0.90±0.02 for coarse-grained), correctly classifying 479/698 cases in fine-grained and 648/698 cases in coarse-grained classification. For fine-grained, UNI(LP) significantly outperformed Prov-GigaPath(LP) (macro recall: 0.73±0.01, patch accuracy: 0.64±0.01) in macro recall ($p=0.020$) and showed a trend towards higher patch accuracy ($p=0.053$). As with the Local dataset, ResNet-RS 50(LP) consistently underperformed (macro recall: 0.29±0.04, patch accuracy: 0.13±0.04) and was excluded from the graphs.

Comparing training conditions within models, UNI's LP condition surpassed FT in both macro recall (p<0.01) and patch accuracy (p=0.011). For $ViT_{IN}$, while LP showed higher macro recall and FT showed higher patch accuracy, the differences were not statistically significant (p=0.78 and p=0.24, respectively), due to inconsistent trends across folds despite differences in mean values. For CTransPath, LP achieved higher macro recall than FT (p=0.037), while FT showed higher patch accuracy than LP (p=0.039).

The EBRAINS confusion matrix (Fig. 2d) revealed distinctive error patterns, with model predictions (G, A, O, M, L) shown against molecular subtypes as ground truth. For *IDH*-wildtype (IDH-) cases, FT showed slightly higher recall than LP (0.76±0.03 vs 0.73±0.07, p=0.431). This was particularly evident in reclassified cases: FT correctly identified more Anaplastic astrocytoma (AA, *IDH*-wildtype) cases as glioblastoma (30/47 vs 24/47) and Diffuse astrocytoma (DA, *IDH*-wildtype) cases (8/19 vs 5/19) compared to LP, which frequently misclassified them as astrocytoma (AA: 6/47 vs 19/47; DA: 4/19 vs 14/19). For *IDH*-mutant (IDH+) astrocytomas, LP significantly outperformed FT (0.59±0.06 vs 0.34±0.11, p<0.001). The difference was most pronounced in DA, IDH+, where LP correctly classified 58/70 cases as astrocytoma, while FT split classifications between astrocytoma (34/70) and oligodendroglioma (24/70). For oligodendrogliomas (*IDH*-mutant and 1p/19q co-deleted), FT showed higher recall than LP (0.59±0.13 vs 0.52±0.05, p=0.167). FT demonstrated better discrimination, particularly in Oligodendroglioma (O) cases (65/85 correct), while LP showed more frequent misclassification as astrocytoma (O: 28/85, Anaplastic oligodendroglioma (AO): 24/91). Additionally, in terms of PCNSL classification, while both conditions showed similar PCNSL sensitivity (57 correct cases each), LP demonstrated more selective prediction (57/83 vs 57/102 predicted cases).

## Effect of Patch Count per Case

We evaluated model performance across varying numbers of training patches per case (500, 100, 25, and 10) on both Local and EBRAINS datasets. The detailed results for 100, 25, and 10 patches are shown (Supplemental Tables 2, 3, and 4, respectively). In the validation folds of the Local dataset, all models showed monotonic improvements in both macro recall and correct patch counts as patch numbers increased (Fig. 3a). Class-wise accuracy analysis revealed increasing trends for G, A, M, and L classes, while the A class showed overall improvement despite some instability (Fig. 3b).

On the EBRAINS dataset, while macro recall generally showed similar monotonic improvements (Fig. 3c), UNI(FT) and Prov-GigaPath(LP) exhibited performance degradation at 100 patches/case and 500 patches/case. Conversely, overall accuracy and patch-wise accuracy showed slight decreases with increased patch numbers across multiple conditions, including UNI models. Class-specific sensitivity analysis revealed consistent improvements with increased patch numbers for G, M, and L classes across all conditions, while A and O classes showed model-dependent variations (Fig. 3d). Notably, foundation models' LP conditions (Prov-GigaPath, UNI, CTransPath) demonstrated distinct trajectories in A recall compared to other conditions.

Quantitative analysis of cumulative performance differences from the 10 patches/case baseline (Fig. 3e) revealed model-dependent responses to increased patch counts. On the Local dataset, foundation models showed modest gains (Prov-GigaPath(LP): +0.056, UNI(LP): +0.044 in macro recall from 10→500 patches), while models without pathology-specific pretraining demonstrated substantially larger improvements (ViT-L(RI): +0.455, ViT-L$_{IN}$(LP): +0.239). On the EBRAINS dataset, foundation models exhibited minimal improvements or performance degradation (Prov-GigaPath(LP): -0.014 in macro recall, -0.035 in patch accuracy; UNI(FT): -0.008 in macro recall), while conventional models consistently benefited from additional patches (ViT-L(RI): +0.134 in macro recall, +0.302 in patch accuracy).

UNI(LP) with 10 patches/case demonstrated superior performance on the EBRAINS dataset, significantly outperforming both ViT-L(FT) and CTransPath(FT) with 500 patches/case in fine-grained classification (0.743±0.008 vs 0.636±0.024 vs 0.678±0.011, p<0.001 and p<0.001) and coarse-grained classification (0.927±0.006 vs 0.866±0.022 vs 0.880±0.015, p<0.001 and p<0.001). On the Local dataset, UNI(LP) with 10 patches/case showed comparable performance in both fine-grained classification (0.840±0.081 vs 0.833±0.084 vs 0.879±0.091, p=0.757 and p=0.209) and coarse-grained classification (0.961±0.039 vs 0.982±0.015 vs 0.989±0.018, p=0.089 and p=0.061).

## Cross-dataset Comparison of Model Decisions

We compared model decision patterns between the Local dataset (Fig. 4) and the EBRAINS dataset (Fig. 5). In both datasets, glioblastoma cases with small hyperchromatic nuclear patterns were sometimes misclassified as metastatic tumors or PCNSL. In astrocytoma and oligodendroglioma cases, classification consistency decreased in regions where the presence or absence of perinuclear halos was ambiguous. For PCNSL, both datasets showed accurate classification in areas with distinct perivascular infiltration, while regions with diffuse parenchymal infiltration were often misclassified as GBM.

Regarding staining characteristics, the EBRAINS dataset contained a notable proportion of cases showing predominantly blue staining patterns. While this variation in staining likely reflects differences in specimen preparation processes between institutions, no direct correlation was observed between staining characteristics and model classification accuracy.

## Analysis of GradCAM Patterns

We analyzed GradCAM patterns by selecting cases with high (>0.9) and low (<0.5) patch-level accuracy per case (Fig. 4 and Fig. 5). The analysis revealed that models consistently detected key diagnostic features: in glioblastoma, strong GradCAM signals highlighted bizarre nuclei and microvascular proliferation, while necrotic regions showed limited response. In PCNSL, distinct signals were observed in perivascular infiltration patterns, a crucial diagnostic feature. Diffusely infiltrating tumors (astrocytoma and oligodendroglioma) showed broader GradCAM signal distribution across tumor regions, contrasting with the more localized signals in solid tumors like metastases. These patterns suggest that the models learned to recognize histologically relevant features for classification.

## Feature Space Analysis

We analyzed the feature space of each model through UMAP visualization. The analysis used data from one fold of the Local dataset (train+val, 10 patches/case) and the EBRAINS dataset (5 patches/case), comparing distributions across datasets and subtypes under four model conditions (Original/Trained ViT$_{IN}$ and Original/Trained UNI) (Fig 6).

Original ViT$_{IN}$, trained solely on natural images, showed clear distributional differences between Local and EBRAINS datasets but demonstrated poor separation between tumor types (Fig 6a). In contrast, Original UNI, pretrained on histopathological images, exhibited tumor type separation across both datasets even without local training (Fig 6c). Notably, Local dataset samples formed distinct clusters at the periphery of the feature space, while EBRAINS samples tended to concentrate toward the center.

After training on the Local dataset, ViT-L$_{IN}$ showed improved separation between tumor types (Fig 6b), while UNI demonstrated even more pronounced cluster formation within the Local dataset (Fig 6d). Detailed analysis of glioma cases in the EBRAINS dataset revealed that the UNI model showed distribution patterns that appeared to reflect morphological grades rather than IDH status. Additionally, analysis of metastatic brain tumors by primary site demonstrated clustering tendencies in all models after local training. These changes in feature distribution patterns aligned with the models' discriminative performance.

## Web-based Diagnosis Support System

As a demonstration of practical applications, we developed an interactive web-based system for pathological image analysis. The system performs real-time classification with microscope camera integration, visualizes decision processes through adjustable-transparency GradCAM, and enables exploration of feature spaces by projecting analyzed images onto UMAP visualization of dataset patches. While the system is currently under development and does not support whole slide image analysis, interested researchers may contact the authors for access to the demonstration.

# Discussion

In this study, we described a notable discovery in the transfer learning of foundation models, in which both UNI and Prov-GigaPath demonstrated robust classification performance with as few as 10 patches per case, achieving equivalent or superior performance to conventional approaches such as ImageNet-pretrained ViT and CTransPath that required 50 times more patches. This result not only validates the potential impact of foundation models specifically designed for pathological imaging but also suggests possibilities for diagnostic support in situations where only limited specimens are available, such as biopsies.

The divergent performance trajectories with increased patch counts suggest that foundation models reach information saturation remarkably early, while conventional models continue to benefit from additional data. This phenomenon may represent a form of overfitting specific to foundation models, where they rapidly adapt to institution-specific patterns at the expense of generalization capability. The pronounced effect in external validation further supports this interpretation, highlighting the importance of carefully calibrating data volume when deploying these powerful models in clinical settings.

Furthermore, particularly noteworthy is that linear probing outperformed fine-tuning in external datasets, and while accuracy improved with increasing patch count in the Local dataset, this trend was either not observed or showed inverse correlation in the external dataset. These findings prompt a fundamental reconsideration of conventional approaches to pathological image analysis.

These unexpected results suggest a paradigm shift from "training encoders on extensive pathological data" to "querying pre-trained encoders with labeled datasets." Specifically, this paradigm shift positions dataset design as a means of accessing the knowledge embedded in foundation models through thin datasets and non-destructive learning approaches such as linear probing. This approach signifies a transition from collecting large-scale pathological data to efficiently leveraging pre-trained feature representations.

In fact, the performance degradation of fine-tuning compared to linear probing in external datasets is similar to "catastrophic forgetting" [23], a phenomenon known in neural networks where models lose their pre-trained generalization capabilities while adapting to new tasks. Various parameter-efficient methods have been developed to address this issue: EWC (Elastic Weight Consolidation) [24], LoRA (Low-Rank Adaptation) [25], and Bottleneck Adapter [26]. Indeed, a comprehensive examination of histopathological foundation model adaptation strategies [27] demonstrated that while "full fine-tuning" (updating all parameters of the encoder and fully connected layers) and linear probing showed competitive performance within the same dataset, PEFT (Parameter-Efficient Fine-Tuning) using LoRA exhibited superior generalization capabilities. Our results suggest that similar phenomena occur in pathology image foundation models, and the introduction of parameter-efficient methods such as EWC and Bottleneck Adapter, alongside the already proven effectiveness of LoRA, could be promising solutions.

In addition to considering these methodologies, the selection of foundation models themselves is an important consideration. While UNI slightly outperformed Prov-GigaPath in our study, the latter has larger model scale and generally shows superior performance across multiple benchmarks [17,28]. However, this superiority isn't uniform, with task-specific variations reported in the literature, and our results fall within this expected range of variation. While the organ-level distribution in foundation models' pre-training datasets is public, detailed histological type information remains unavailable, and these distributional differences likely determine each model's characteristics. Therefore, model selection should consider not only model scale but also careful examination of published benchmark results and validation performance on actual test data.

A more practical challenge revealed by UMAP projection analysis is the clear distributional divergence between Local and EBRAINS data, likely reflecting inter-institutional staining variations. To address this issue, training methods that account for staining diversity should be considered [29]. As a simple approach, color normalization and other image preprocessing techniques [30] could overcome this challenge. Of particular concern for practical implementation is the low accuracy in fine-grained classification of Astrocytoma and Oligodendroglioma cases. As shown in the confusion matrix, frequent misclassification occurred between these tumor types. This issue is underpinned by the challenge of limited case numbers. Indeed, the small number of these cases in our Local dataset reflects the actual distribution reported in the Japan Neurosurgical Society's comprehensive survey [3], highlighting the inherent constraint of single-institution data collection over a finite period. While addressing this issue requires larger-scale multi-institutional data collection and domain adaptation techniques to enhance generalization capability, developing efficient learning methods for such rare cases is equally important.

Despite these technical challenges, the high accuracy achieved in coarse-grained classification demonstrates strong potential for clinical AI system applications, particularly in intraoperative rapid diagnosis. The differentiation among Glioma, PCNSL, and brain metastases directly influences intraoperative decision-making [31,32], and notably, no rapid diagnostic models specifically designed for these three entities have been previously reported [33–35]. Our web-based diagnostic support system implements real-time classification display, GradCAM visualization, and morphological feature relativization, offering potential benefits for both diagnostic support and education. While the current ROI-based analysis constraint may not significantly impact microscope camera-connected diagnostic support systems, we are investigating whole slide image analysis through Multiple Instance Learning [36] and attention-based implementations [37] for more comprehensive analysis.

## Conclusion

This study has illuminated optimal application strategies for foundation models in pathological image classification while demonstrating a paradigm shift in dataset design. These insights contribute to the theoretical foundation for developing more efficient and reliable medical AI systems.

## Acknowledgements

This work was supported by the grant from Japan Society for the Promotion of Science (JSPS) (S.T. 24H00037) and Japan Science and Technology Agency(JST) BOOST (K.E. JPMJBS2426). We are grateful to CHOWA GIKEN Corporation and Mr. Fuminori Kamakura (NIPPON TELEGRAPH AND TELEPHONE EAST CORPORATION) for their assistance.

## Author contributions

K.E. designed and executed the entire study, including machine learning implementation and manuscript preparation. Y.O. provided pathological annotations and supervised the study methodology. Z.T. provided pathological annotations. K.S., H.M., T.S., and S.Y. contributed clinical cases for the Local dataset. T.O. provided manuscript review. W.L. and M.T. provided valuable feedback. S.T. provided administrative supervision.

## Methods

### Local Datasets

For this study, we retrospectively collected 254 cases from the Department of Cancer Pathology, Hokkaido University, between 2016 and 2023. This included 249 brain tumor cases across five tumor types: 66 cases of Glioblastoma, *IDH*-wild type (GBM), 26 cases of Astrocytoma, *IDH*-mutant (A), 24 cases of Oligodendroglioma, *IDH*-mutant, 1p19q co-deleted (O), 52 cases of PCNSL (L), and 81 cases of metastatic tumors (M). For background tissue, we obtained non-tumor brain specimens from 5 autopsy cases (Fig. 1b).

Glioma classification was based on WHO 2021 criteria, with *IDH* status evaluated through immunohistochemical and genetic testing. All diagnoses were made by experienced pathologists in our department. While cases from 2021 onwards followed WHO 5th edition criteria, earlier cases were reclassified according to current diagnostic criteria based on *IDH* and 1p19q status. For quality control, we excluded cases with extremely small specimen volumes, cases without *IDH* genetic testing, and special cases such as those with heterogeneity in 1p19q deletion.

For specimen preparation, permanent slides were created from formalin-fixed paraffin-embedded tissues and stained with H&E. Slides were digitized using Hamamatsu NanoZoomer XR and S210 scanners to obtain whole slide images (WSI). Using NDP.view2 at 20x magnification (approximately 440 nm/pixel), tumor regions were visually selected and then divided into patches of 512x512 pixels or larger (Fig. 1a). For tumor classes, patches with white areas (pixel values 230-255) exceeding 40% were removed, while this processing was intentionally omitted for background tissue class to allow recognition of non-tumor areas as white matter. The resulting patch count distribution varied across cases (Fig. 1c).

For image preprocessing, we applied the same normalization as ImageNet [21] (MEAN = (0.485, 0.456, 0.406), STD = (0.229, 0.224, 0.225)). Data augmentation was performed using the Albumentations library [38], including geometric transformations, brightness and color adjustments, noise simulation, and structural modifications. The input size was set to 224×224 pixels (approximately 1µm/pixel).

### Models

For image feature extraction, we compared four encoders: the 2024 foundation models UNI [13] and Prov-GigaPath [15], the conventional baseline model CTransPath [18], and ViT-L [14] and ResNet-RS 50 [39]. Prov-GigaPath is based on ViT-Giant, while UNI is based on ViT-L, both using pretrained weights publicly available on huggingface [28,40]. For ViT-L, we used the same ViT-L/16 model as UNI, with parameters pretrained on ImageNet-21k, obtained through the timm library [41] (internally identified as vit_large_patch16_224.augreg_in21k_ft_in1k). ResNet-RS 50's ImageNet-pretrained parameters were pretrained on ImageNet-1k, also obtained through the timm library (internally identified as resnetrs50.tf_in1). These models are denoted with the suffix IN(ImageNet), such as ViT-L$_{IN}$ and ResNet-RS 50$_{IN}$.

We define three training conditions, indicated by suffixes: Linear probing (LP), where the pretrained encoder parameters are frozen and only the fully connected layers are updated; Fine-tuning (FT), where both pretrained encoder and fully connected layer parameters are simultaneously updated; and Random init (RI), where the encoder is randomly initialized without pretraining and all parameters are updated during training(Fig. 1a).

In this study, we compare ten conditions: Prov-GigaPath(LP), UNI(FT), UNI(LP), CTransPath(FT), CTransPath(LP), ViT-L$_{IN}$(FT), ViT-L$_{IN}$(LP), ViT-L(RI), ResNet-RS 50$_{IN}$(FT), and ResNet-RS 50$_{IN}$(LP).

## Training

We conducted training and evaluation using 5-fold cross-validation, with varying patch limits per case (10, 25, 100, and 500 patches). For cases exceeding the patch limit, excess images were randomly excluded; for cases below the limit, images were used repeatedly during training. Training employed a batch size of 50 with the Adam optimizer[42]. Learning rates remained fixed throughout training: 0.001 for LP, 0.00001 for ViT-based models (UNI, CTransPath, ViT-L), and 0.0001 for CNN models. All encoders were connected to a single fully connected layer with size $F \times C$ (where $F$ is the encoder's output feature dimension and $C$=6). For coarse-grained classification, probabilities of G, A, and O classes were summed to represent the glioma class. The output was followed by softmax activation, and the model was trained using cross-entropy loss. Training was conducted for 30 epochs across all conditions, with early stopping applied when validation loss showed no improvement over 5 consecutive epochs.

## Evaluation metrics

For case classification, we applied majority voting across patches from each case by selecting the most frequently predicted class. We calculated both patch-wise and macro-averaged metrics:

$$Patch\ Accuracy = \frac{Correctly\ classified\ patches}{Total\ patches}$$

$$Overall\ Accuracy = \frac{TP + TN}{TP + TN + FP + FN}$$

$$Macro\ Metric = \frac{1}{K}\sum Metric_i$$

where $K$ is the number of classes, $i$ represents each class, and $Metric_i$ is one of:

$$Recall = \frac{TP}{TP + FN}$$

$$Precision = \frac{TP}{TP + FP}$$

$$F1\ Score = \frac{2 \times (Precision \times Recall)}{Precision + Recall}$$

Here *TP* (true positive), *TN* (true negative), *FP* (false positive), and *FN* (false negative) denote correctly and incorrectly classified samples for each class. For the Local dataset, Overall Accuracy was calculated by combining results from all validation folds. For the EBRAINS dataset, Overall Accuracy was calculated using ensemble predictions from all folds, where predictions were averaged across the softmax outputs of all fold models. Given the imbalanced distribution across datasets and diagnoses, we placed emphasis on macro-averaged recall while also considering patch-wise accuracy and overall accuracy for comprehensive evaluation. Statistical significance between models was evaluated using paired t-tests on validation accuracies across 5-fold cross-validation (p < 0.05). 95% confidence intervals were calculated as mean ± 1.96×SE, where SE is the standard error of the mean across folds.

## External dataset validation

We collected WSI data from the EBRAINS[22] database corresponding to diagnoses in our internal dataset. For each diagnostic category, we aimed to maintain a balanced representation while considering data quality and computational resources:

- For the Glioblastoma (G) class, we included cases of "Glioblastoma, *IDH*-wildtype" (199/474), "Anaplastic astrocytoma, *IDH*-wildtype" (47/47), and "Diffuse astrocytoma, *IDH*-wildtype" (19/19), totaling 265 cases. These latter two categories are now classified as "Molecular glioblastoma" under WHO 5th edition criteria when they demonstrate specific molecular features.

- For the Glioblastoma (G) class, we included cases of "Glioblastoma, *IDH*-wildtype" (199 cases from 474 available cases, as 200 cases were downloaded due to technical constraints in database access and one case was excluded due to data corruption), "Anaplastic astrocytoma, *IDH*-wildtype" (47/47), and "Diffuse astrocytoma, *IDH*-wildtype" (19/19), totaling 265 cases. These latter two categories are now classified as "Molecular glioblastoma" under WHO 5th edition criteria when they demonstrate specific molecular features.

- For the Astrocytoma (A) class, we included "Diffuse Astrocytoma, *IDH*-mutant" (70/70), "Anaplastic astrocytoma, *IDH*-mutant" (47/47), and "Glioblastoma, *IDH*-mutant" (34/34), totaling 151 cases.

- The Oligodendroglioma (O) class comprised "Anaplastic oligodendroglioma, *IDH*-mutant and 1p/19q codeleted" (91/91) and "Oligodendroglioma, *IDH*-mutant and 1p/19q codeleted" (85/85), totaling 176 cases.

- The Brain metastasis (M) class comprised cases from "Metastatic tumours" (47/47).

- The PCNSL (L) class included cases from "Diffuse large B-cell lymphoma of the CNS" (59/60), with one case excluded due to file corruption.

From the "Glioblastoma, *IDH*-wildtype" group, case selection was based on completeness of clinical data and image quality. Image preprocessing and ROI extraction followed the same methodology as our Local dataset, yielding 24.2 ± 14.6 (mean ± SD) patches per case from ROI images averaging 3562.9 ± 886.4 × 2001.4 ± 494.4 (mean ± SD) pixels.

## Analysis of Inference Process

We employed two complementary visualization techniques. First, we utilized GradCAM [43] to visualize the final encoder layers of both ViT-based and CNN models, examining which regions contribute most significantly to their classification decisions. Second, we analyzed the distribution of features in the high-dimensional space by visualizing feature embeddings using UMAP [44]. These features were extracted from each encoder's output after pooling and before input to the fully connected layers.

## Implementation Details

Training was conducted in parallel on two NVIDIA RTX 3090 GPUs (24GB VRAM) using PyTorch 2.3.0 and Python 3.12.4.Ethical approval

This retrospective study was approved by the ethics committee of Hokkaido University Hospital (024-0476), and the requirement for written informed consent was waived. All studies were performed in accordance with relevant guidelines and regulations.

## Data and Code Availability

The implementation code and analysis scripts for this study are publicly available at [https://github.com/endaaman/bt](https://github.com/endaaman/bt). The Local dataset can be obtained by contacting the corresponding author. We are committed to research transparency and will work to facilitate data sharing requests while ensuring compliance with privacy requirements.

## Competing Interests Statement

The authors declare no competing interests.

# Tables

## Table 1: Performance of Models Trained with 500 Patches/Case on Local and EBRAINS Datasets

|  |  | Fine-grained | | | Coarse-grained | | |
| --- | --- | --- | --- | --- | --- | --- | --- |
|  |  | Macro Recall | Patch Acc. | Overall Acc. | Macro Recall | Patch Acc. | Overall Acc. |
| Local | Prov-GigaPath(LP) | 0.88 ± 0.10 | 0.93 ± 0.02 | 234/254 | 0.99 ± 0.01 | 0.97 ± 0.01 | 251/254 |
|  | UNI(FT) | 0.87 ± 0.07 | 0.94 ± 0.01 | 231/254 | 0.99 ± 0.02 | 0.97 ± 0.01 | 250/254 |
|  | UNI(LP) | 0.88 ± 0.09 | 0.93 ± 0.03 | 234/254 | 0.99 ± 0.02 | 0.97 ± 0.01 | 250/254 |
|  | CTransPath(FT) | 0.88 ± 0.09 | 0.92 ± 0.02 | 233/254 | 0.99 ± 0.02 | 0.96 ± 0.01 | 250/254 |
|  | CTransPath(LP) | 0.87 ± 0.06 | 0.90 ± 0.03 | 230/254 | 0.98 ± 0.02 | 0.95 ± 0.01 | 248/254 |
|  | ViT-L$_{IN}$(FT) | 0.83 ± 0.08 | 0.91 ± 0.02 | 225/254 | 0.98 ± 0.02 | 0.95 ± 0.02 | 248/254 |
|  | ViT-L$_{IN}$(LP) | 0.83 ± 0.05 | 0.86 ± 0.03 | 217/254 | 0.94 ± 0.02 | 0.91 ± 0.02 | 235/254 |
|  | ViT-L(RI) | 0.83 ± 0.04 | 0.86 ± 0.02 | 215/254 | 0.93 ± 0.04 | 0.91 ± 0.02 | 233/254 |
|  | ResNet-RS 50$_{IN}$(FT) | 0.85 ± 0.10 | 0.91 ± 0.02 | 229/254 | 0.99 ± 0.01 | 0.96 ± 0.01 | 250/254 |
|  | ResNet-RS 50$_{IN}$(LP) | 0.36 ± 0.07 | 0.38 ± 0.09 | 57/254 | 0.50 ± 0.09 | 0.52 ± 0.06 | 126/254 |
| EBRAINS | Prov-GigaPath(LP) | 0.73 ± 0.01 | 0.64 ± 0.01 | 474/698 | 0.92 ± 0.00 | 0.88 ± 0.01 | 635/698 |
|  | UNI(FT) | 0.71 ± 0.02 | 0.64 ± 0.02 | 470/698 | 0.91 ± 0.02 | 0.87 ± 0.04 | 625/698 |
|  | UNI(LP) | 0.75 ± 0.01 | 0.65 ± 0.02 | 479/698 | 0.94 ± 0.01 | 0.90 ± 0.02 | 648/698 |
|  | CTransPath(FT) | 0.68 ± 0.01 | 0.59 ± 0.03 | 437/698 | 0.90 ± 0.01 | 0.86 ± 0.03 | 626/698 |
|  | CTransPath(LP) | 0.69 ± 0.01 | 0.59 ± 0.01 | 446/698 | 0.90 ± 0.01 | 0.85 ± 0.02 | 632/698 |
|  | ViT-L$_{IN}$(FT) | 0.64 ± 0.02 | 0.56 ± 0.04 | 422/698 | 0.87 ± 0.02 | 0.81 ± 0.05 | 602/698 |
|  | ViT-L$_{IN}$(LP) | 0.64 ± 0.02 | 0.53 ± 0.04 | 407/698 | 0.84 ± 0.02 | 0.74 ± 0.05 | 561/698 |
|  | ViT-L(RI) | 0.54 ± 0.04 | 0.44 ± 0.05 | 336/698 | 0.76 ± 0.03 | 0.69 ± 0.06 | 537/698 |
|  | ResNet-RS 50$_{IN}$(FT) | 0.62 ± 0.03 | 0.54 ± 0.03 | 416/698 | 0.86 ± 0.01 | 0.79 ± 0.03 | 605/698 |
|  | ResNet-RS 50$_{IN}$(LP) | 0.29 ± 0.04 | 0.13 ± 0.04 | 189/698 | 0.45 ± 0.04 | 0.46 ± 0.16 | 549/698 |

Values shown as mean ± 95% confidence interval from 5-fold cross-validation; Overall Accuracy for Local dataset represents combined validation folds, while EBRAINS represents ensemble predictions.

# Figures

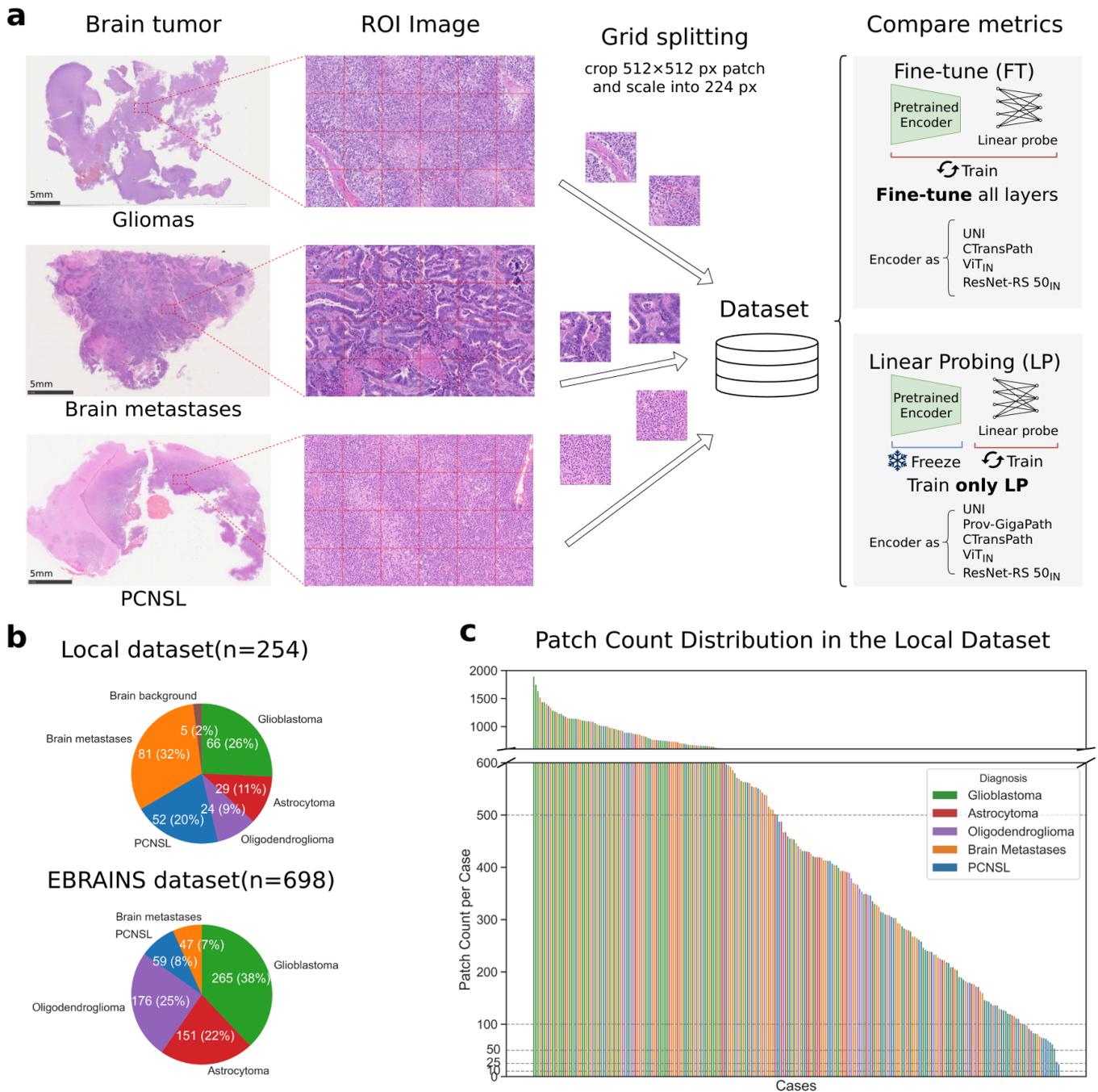

**Fig. 1 Dataset Curation and Model Training Strategy**

a) Study workflow showing ROI selection from hematoxylin and eosin-stained whole slide images and patch extraction (left), and model training approaches (right): fine-tuning (FT) with full parameter updates and linear probing (LP) with frozen encoder weights.

b) Distribution of patch counts in the Local dataset, with horizontal lines indicating the training patch count conditions (10, 25, 100, and 500 patches).

c) Distribution of tumor classes (glioblastoma, astrocytoma, oligodendroglioma, primary CNS lymphoma, and metastatic tumors) shown in pie charts for Local dataset (n=254, upper) and EBRAINS dataset (n=698, lower).

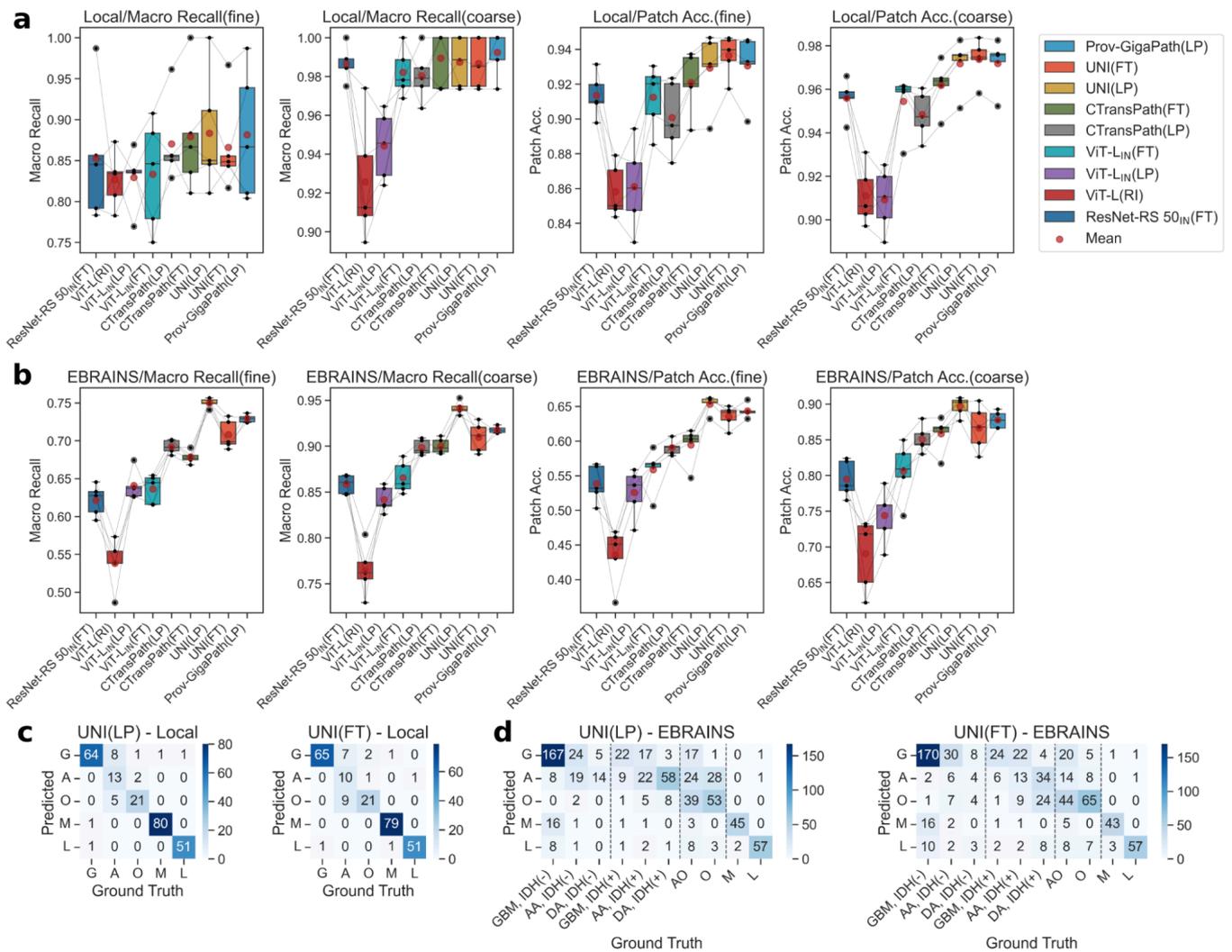

**Fig. 2 Model Performance Evaluation and Classification Analysis**

a) Box plots comparing model performances on the Local dataset validation sets with 500 patches/case. From left to right: macro recall and patch accuracy for both fine-grained and coarse-grained classification tasks. Box plots show quartiles: the box extends from Q1 (25th percentile) to Q3 (75th percentile), with a line at Q2 (median, 50th percentile). Whiskers extend to the min and max values within 1.5 times the interquartile range. Each point represents a validation fold metric, with lines connecting points from the same fold. Red points indicate mean values across folds. ResNet-RS 50(LP) was excluded from panels (a) and (b) due to substantially lower performance (see Table 1).

b) Box plots showing model performances on the EBRAINS dataset, evaluated using models trained with 500 patches/case. Box plot quartile interpretation matches panel (a).

c) Confusion matrices for the Local dataset (left: UNI(LP), right: UNI(FT)), combining results from all validation folds. Classes are glioblastoma (G), astrocytoma (A), oligodendroglioma (O), metastatic tumors (M), and PCNSL (L). d) Confusion matrices for the EBRAINS dataset using ensemble predictions from all folds (left: UNI(LP), right: UNI(FT)). Ground truth columns for glioma classes are expanded into molecular subtypes - G class: GBM, AA, and DA, all IDH(−); A class: GBM, AA, and DA, all IDH(+); O class: AO and O (GBM: Glioblastoma, AA: Anaplastic astrocytoma, DA: Diffuse astrocytoma, AO: Anaplastic oligodendroglioma, O: Oligodendroglioma).

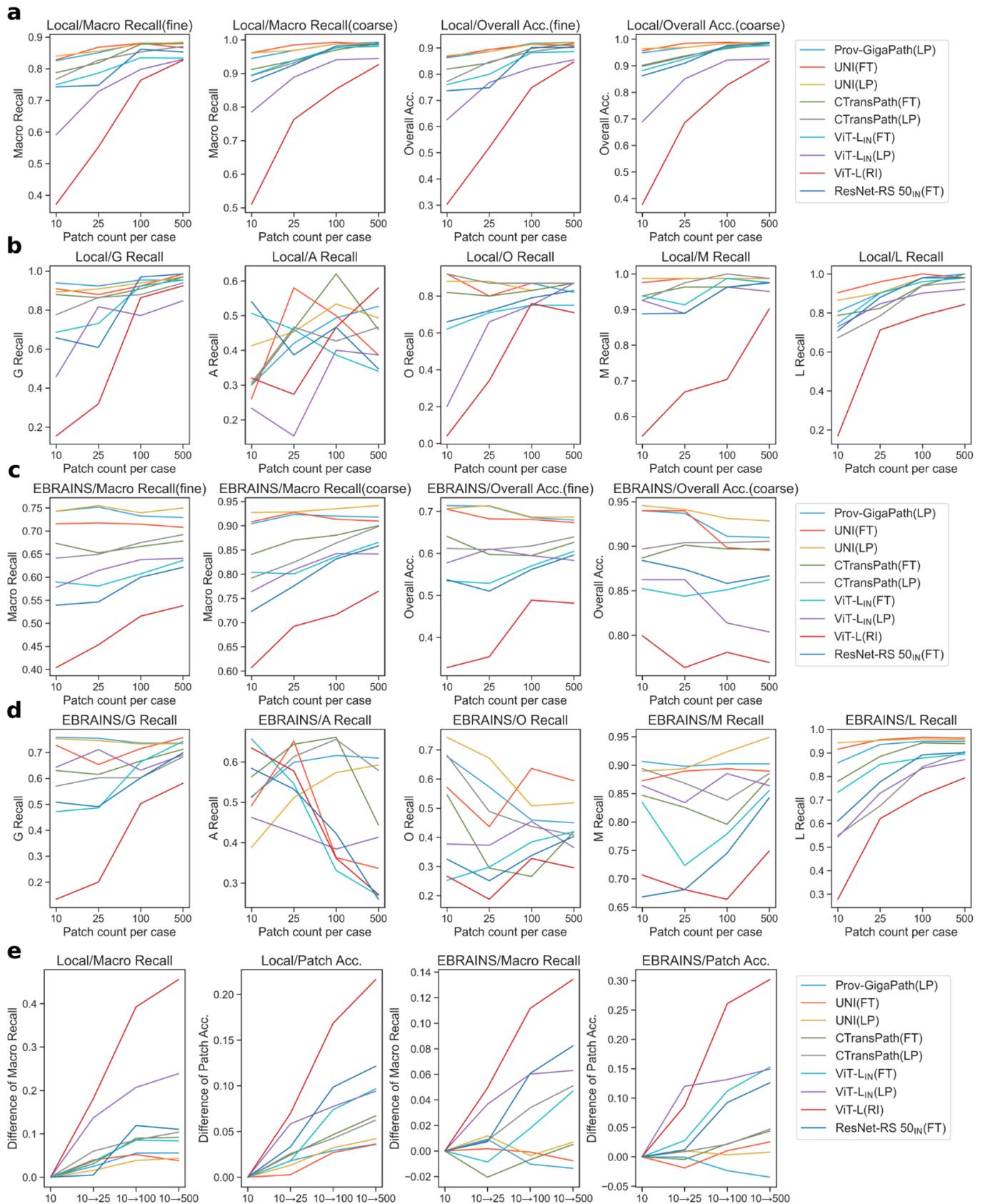

**Fig. 3 Effect of Training Patch Count on Model Performance**

a) Model performance trends on Local dataset validation sets with increasing patch counts (10, 25, 100, and 500 patches/case). From left to right: macro recall and overall accuracy for both fine-grained and coarse-grained classification tasks. Each line represents a different model configuration, with values calculated by pooling results across all validation folds. b) Class-wise recall trends on the Local dataset showing individual performance for glioblastoma (G), astrocytoma (A), oligodendroglioma (O), metastatic tumors (M), and PCNSL (L) across different patch count conditions. c) Model performance trends on the EBRAINS dataset, evaluated using models trained with different patch counts. Metrics and layout match panel (a). d) Class-wise recall trends on the EBRAINS dataset. Layout matches panel (b). e) Cumulative differences from 10 patches/case baseline across increasing patch counts. Plots show Local and EBRAINS datasets results for macro recall and patch accuracy (left to right).

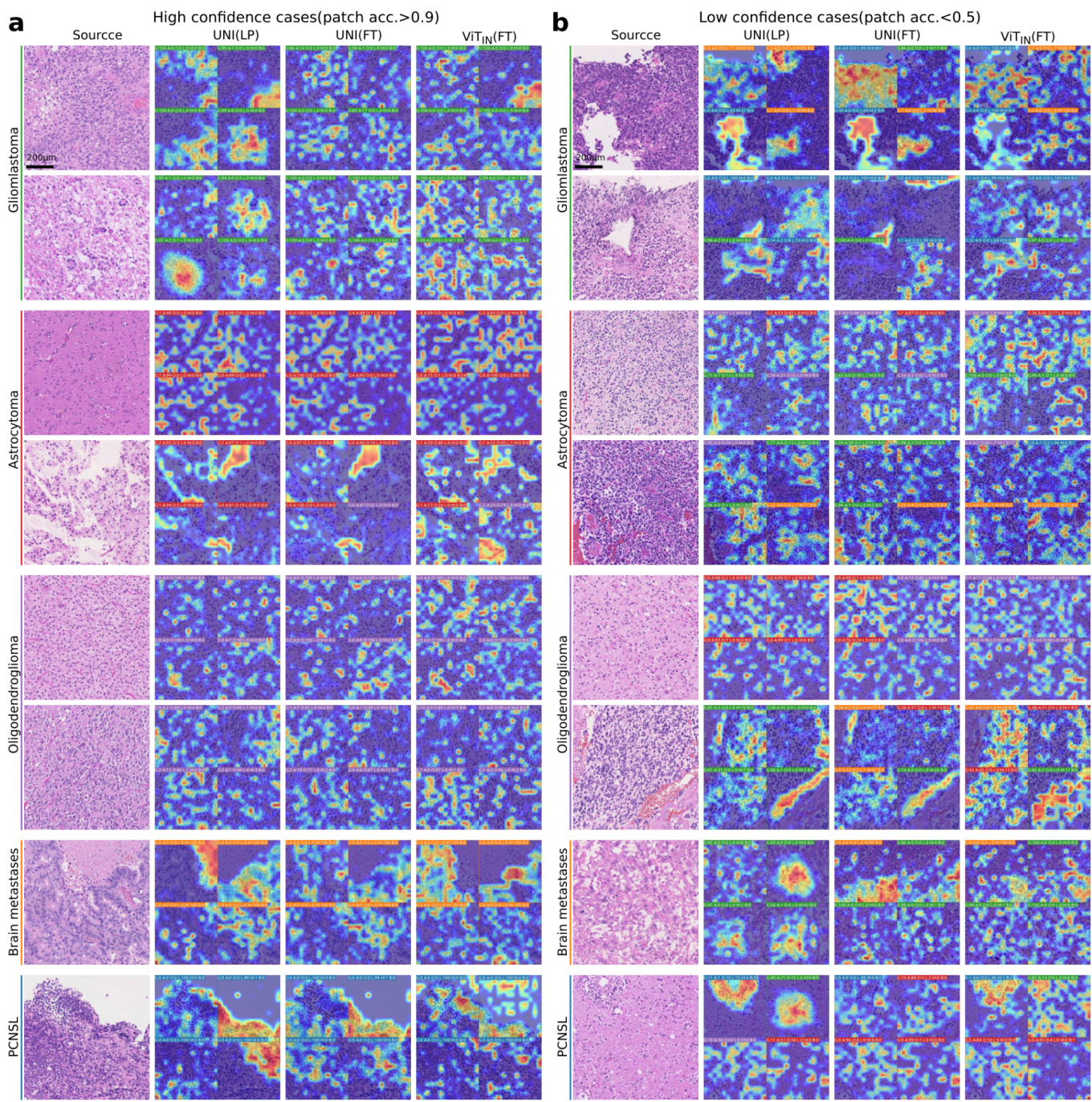

**Fig. 4 Model Performance Analysis Through GradCAM Visualization of Local Dataset Cases**

Representative patches selected from validation sets of Local dataset cases with varying classification performance. Each panel shows original H&E images (leftmost column, with 200 µm scale bar) followed by GradCAM visualization overlays from different models: UNI with linear probing (UNI(LP)), UNI with fine-tuning (UNI(FT)), and ViT-L$_{IN}$ (FT). Each model's visualization is divided into four subpatches, with prediction probabilities shown for each tumor class. a) High-confidence cases (patch accuracy > 0.9). b) Low-confidence cases (patch accuracy < 0.5). GradCAM overlays use heat map coloring where red indicates regions of highest importance for classification.

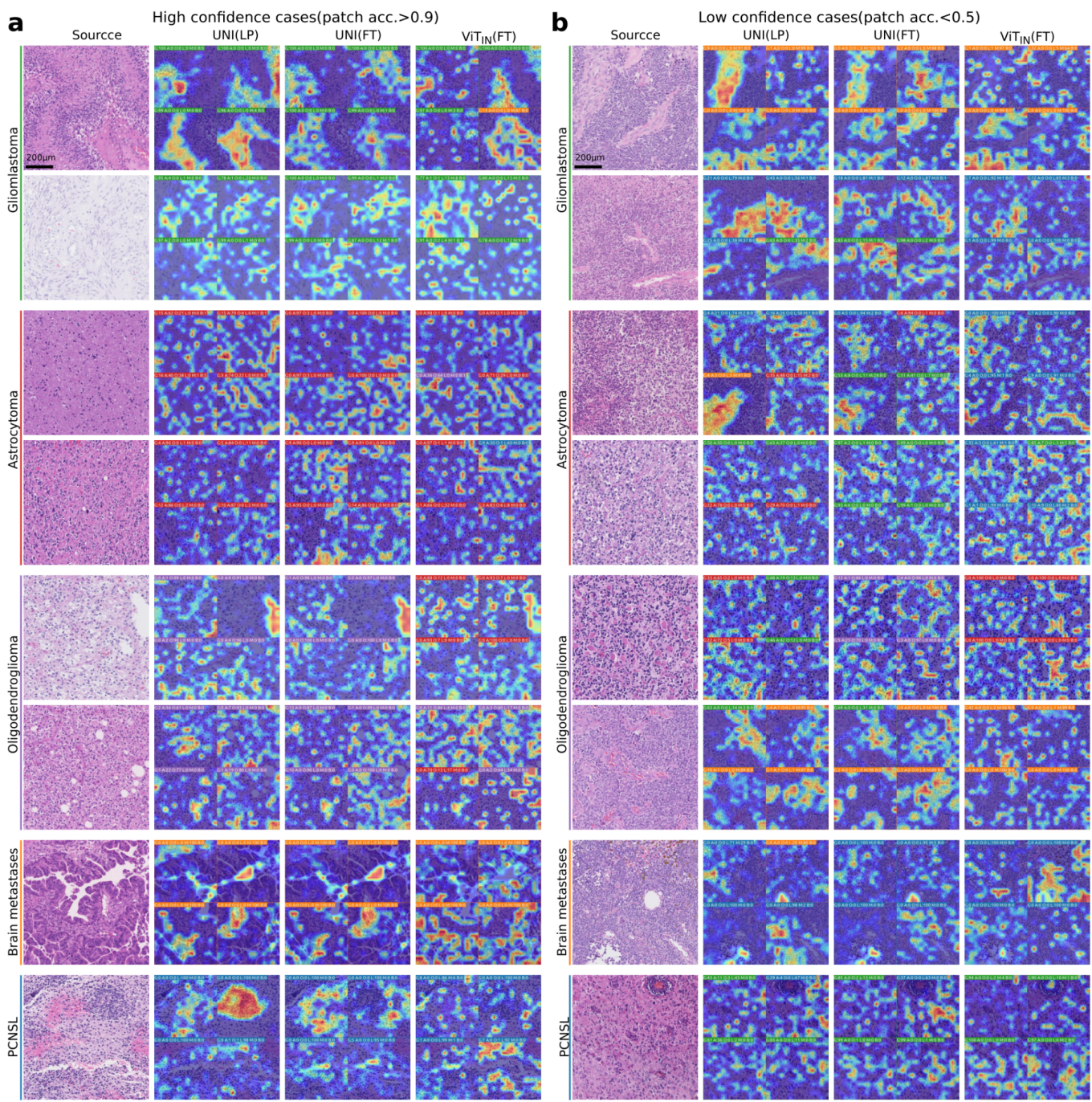

**Fig.5 GradCAM Visualization Analysis of EBRAINS Dataset Cases**

Representative patches selected from EBRAINS dataset cases with varying classification performance, evaluated using a single validation fold. Each panel shows original H&E images (leftmost column, with 200 μm scale bar) followed by GradCAM visualization overlays from different models: UNI with linear probing (UNI(LP)), UNI with fine-tuning (UNI(FT)), and ImageNet-pretrained ViT-L$_{IN}$ (FT). Each model's visualization is divided into four subpatches, with prediction probabilities shown for each tumor class. a) High-confidence cases (patch accuracy > 0.9). b) Low-confidence cases (patch accuracy < 0.5). GradCAM overlays use heat map coloring where red indicates regions of highest importance for classification.

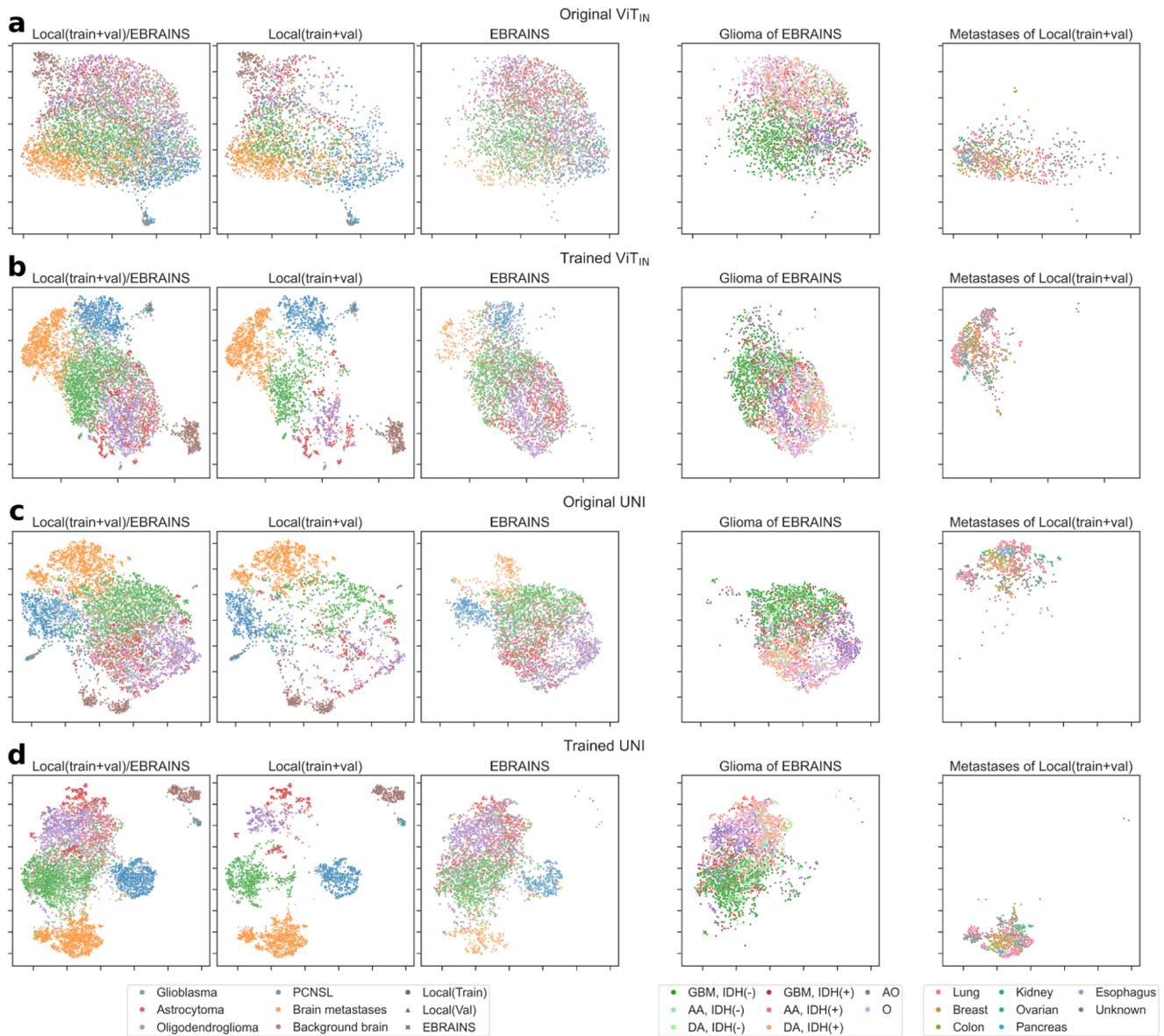

**Fig6: Feature Space Analysis Using UMAP Visualization**

Feature space visualization using UMAP (n_neighbors=70, min_dist=0.5, spread=1.0) comparing the distribution of 1024-dimensional features across datasets and tumor subtypes under four model conditions: a) Original ViT-L$_{IN}$, trained solely on natural images. b) Trained ViT-L$_{IN}$, after training on Local dataset. c) Original UNI, pretrained on pathological images. d) Trained UNI, after training on Local dataset.

From left to right: combined visualization of Local(train+val, 10 patches/case) and EBRAINS (5 patches/case) datasets, Local(train+val) only, EBRAINS only, glioma subtypes from EBRAINS (G class: GBM, AA, and DA, all IDH(−); A class: GBM, AA, and DA, all IDH(+); O class: AO and O; color intensity indicating grade), and metastatic tumors from Local dataset by primary site. Results shown for a single validation fold.

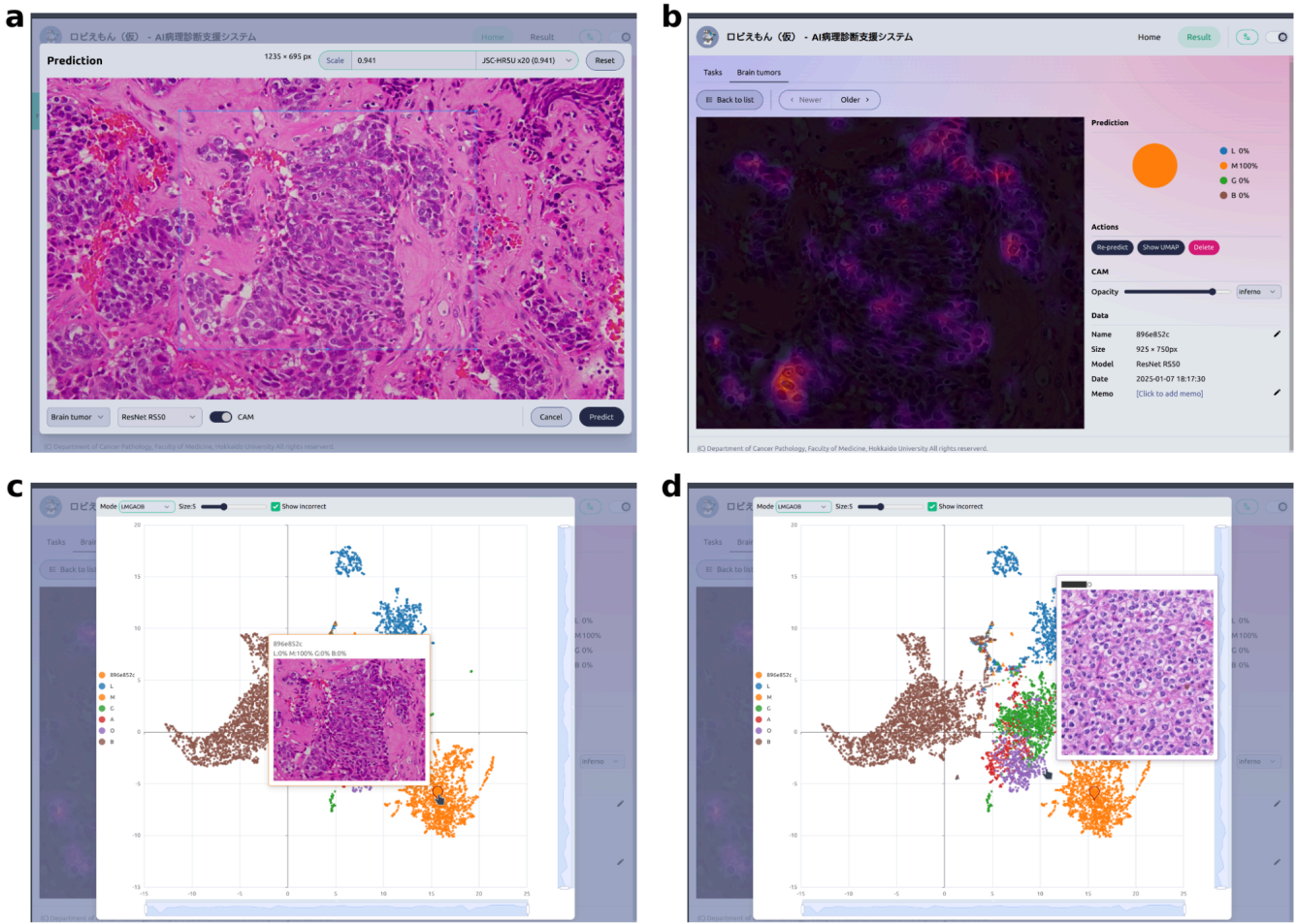

**Fig7: Interactive Web-based System for Real-time Diagnostic Support**

Prototype interface demonstrating real-time diagnostic support capabilities:

a) Analysis configuration interface for images captured through WebRTC-enabled microscope camera integration, providing options for region of interest selection, scale calibration (to maintain consistent pixels-per-micron ratios across different magnifications), model selection, and GradCAM visualization settings.

b) Results display interface featuring class probability distribution as a pie chart and GradCAM overlay with adjustable opacity.

c,d) Interactive feature space visualization using UMAP: c) Real-time projection of analyzed images onto the feature space, with hover functionality showing the corresponding histological image. d) Interactive feature space visualization using UMAP: exploration of training dataset patches through hover interaction, enabling visual comparison between analyzed and training samples.